\documentclass{aa}  
\usepackage{graphicx}
\usepackage{txfonts}
\begin{document}
  \title{A search for water masers toward extrasolar planets}

   \author{V. Minier
              \inst{1,2,3}
          \and C. Lineweaver
          \inst{3,4}}

   \offprints{Vincent Minier }

\institute{Service d'Astrophysique, DAPNIA/DSM/CEA Saclay, 91 191 Gif-sur-Yvette, France $-$ 
            \email{vincent.minier@cea.fr}  
  \and AIM, Unit\'e Mixte de Recherche, CEA$-$CNRS$-$Universit\'e Paris VII, UMR 7158, CEA/Saclay, 91191 Gif-sur-Yvette, France 
  \and School of Physics, University of New South Wales, Sydney NSW 2052, Australia 
  \and Planetary Science Institute RSAA/RSES, Australian National University, Canberra ACT 0200, Australia}

   \date{Received / Accepted}

% \abstract{}{}{}{}{} 
% 5 {} token are mandatory
 
  \abstract
  % context heading (optional)
  % {} leave it empty if necessary  
   {Water is the most common triatomic molecule in the universe and the basis of life on Earth.
             Astrophysical masers have been widely studied in recent years and have been shown 
             to be invaluable probes of the details of the environment in which they are found. 
             Water masers, for instance, are often detected toward low-mass star-forming regions. Doppler 
radial-velocity surveys have detected about 160 exoplanets.}
  % aims heading (mandatory)
   {Observations of water masers from exoplanetary systems would give us a new detailed 
             window through which to explore them.}
  % methods heading (mandatory)
   {We present a search for water masers toward eighteen extrasolar planets using the newly upgraded Australia
             Telescope Compact Array at 12 mm. A sensitivity of $\sim25$~mJy~beam$^{-1}$ and an angular resolution of $\sim10''$
             were achieved at 22.235 GHz. }
  % results heading (mandatory)
   {No maser lines are clearly observed. }
  % conclusions heading (optional), leave it empty if necessary 
   {}

   
   \maketitle
%
%________________________________________________________________

\section{Introduction}

Water is the most common triatomic molecule in the Universe and the basis of life on Earth. Doppler 
radial-velocity surveys have detected about 160 planets\footnote{As reported in
http://www.obspm.fr/planets in June 2005} orbiting nearby solar-type stars including 
18 multiple planet systems. The proximity and availability of water on or near these exoplanets is an 
important piece of our emerging picture of how our Solar System compares to these newly detected 
planetary systems, and more speculatively what the prospects for water-based life are.

Cosmic thermal water emission is not easily detectable with ground-based telescopes
because the large amount of vapour in the lower atmosphere of the Earth contributes
to the emission itself and absorbs radiations in the sub-millimetre and infrared parts
of the electromagnetic spectrum, where water spectral lines are
largely present (e.g. Boonman et al. \cite{boonman03}; Deguchi \& Nguyen-Q-Rieu \cite{deguchi90}). 
However, non-thermal water emission is observed from Earth: water
vapour masers (e.g. Cheung et al. \cite{cheung69}; Waters et al. \cite{waters80}) are very intense and widespread cosmic phenomena that have been
detected toward star-forming regions (Genzel \& Downes \cite{genzel79}), late
M-type stars (Dickinson \cite{dickinson76}) and Active Galactic Nuclei (Claussen et
al. \cite{claussen84}).  They were detected  at 22.235 GHz (Cheung et
al. \cite{cheung69}), 183.309 GHz (Cernicharo et al. \cite{cernicharo90}), 321~GHz and 325~GHz (Menten et al. \cite{menten90a}, \cite{menten90b}).  

Recently, detections of various types of water emission in the Solar
System have been reported. Water abundances were measured in the
atmosphere of giant planets with the {\it ISO} and {\it SWAS} space
observatories (e.g. Bergin et al. \cite{bergin00}) and in comets with the {\it ODIN} 
satellite (Lecacheux et al. \cite{lecacheux03}). Thermal water
emission and 22-GHz water masers were also seen coming from the impact
on Jupiter induced by the Shoemaker-Levy comet collision (Bjoraker et
al. \cite{bjoraker96}; Cosmovici et al. \cite{cosmovici96}). Finally, water
absorption features were identified toward the sunspot umbrae and might originate on
the Sun (Wallace et al. \cite{wallace95}).

Traces of water emission have also been
reported near solar-type star systems. Water masers were for instance
imaged in NGC~2071 where they might trace a protoplanetary disk around
a 1-M$_{\odot}$ protostar (Torrelles et al. \cite{torrelles98}). Evidence for water
emission was discovered toward early M-type giant stars (Tsuji et
al. \cite{tsuji97}). Water masers also arise in the circumstellar envelopes of
more evolved giant stars such as Mira and AGB stars (e.g. Takaba et
al. \cite{takaba01}) and continue to glare in post-AGB stars and young
planetary nebulae (e.g. Engels \cite{engels02}). This demonstrates that
H$_2$O can be present in the environment of a wide range of stars in
the HR diagram. 

In 2002, possible detections of water masers toward extrasolar planets were presented by 
Cosmovici et al. (\cite{cosmovici02}) during the Second European Workshop on Exo/Astrobiology in Graz, Austria
although the results have remained unpublished\footnote{This possible discovery was announced through
New Scientist 175(2361), 22}.
The maser lines were possibly detected toward Epsilon Eridani, Upsilon Andromedae and 47 Ursa Majoris at a level
of a few 100 mJy (Cosmovici, priv. comm.). 
However,  null results have also been announced for five exoplanet host stars down to a sensitivity of 14-64 mJy (Greenhill, 
IAU circular 7985), of which Upsilon Andromedae and Epsilon Eridani had been reported to emit a possible water maser. 
Non-detections were also reported by Butler et al. (\cite{butler02}) after observing Epsilon Eridani, Upsilon Andromedae and 
47 Ursa Majoris down to a sensitivity of 2.6-10.2 mJy with the VLA.

In this letter, we present the results of a search for water masers toward 18 extrasolar planets. 
The objective was to study 
whether water masers can form in the near environment (e.g. planetary atmosphere, circumstellar environment or comet clouds) of 
extrasolar planets. The targets were selected to be observable from the Southern latitudes out of $\sim150$ known exoplanets available at the date of the observations. The high sensitivity and high angular resolution
search was conducted with the Australia Telescope Compact Array, that was 
upgraded in April 2003 to operate at 12 mm.
Observations of water masers from exoplanetary systems would give us a new detailed window through which to 
explore them. Details extractable from maser detection include answers to the following questions: what is 
the velocity of the masing source (e.g. planets or comets)? Which part of the planetary system is compatible 
with the column densities of water, a pumping mechanism and a lack of collisional thermalization that would 
otherwise quench the maser? By combining observations and models, the physical conditions of the exoplanetary 
systems may be probed. 

\section{Conditions for detecting water masers toward extrasolar planets}

The mechanisms that might generate 22-GHz water maser emission in extrasolar planets include (but are not limited to) 
cometary impacts in atmospheres of giant planets, particularly in younger stellar systems in which much more massive 
and frequent impacts are expected (e.g. Chyba \cite{chyba90}; Shoemaker \cite{shoemaker83}). Whether the required conditions (column density, 
temperature, 
velocity coherence, path length, pumping) are present to produce detectable water masers is unknown. 

Water maser action needs a large column density of water vapour and a pumping mechanism
to ensure the inversion of the population levels $6_{16}-5_{23}$. 
22-GHz masers are probably pumped by collisions with H$_2$, although
radiative pumping is also proposed to explain masers at higher
frequencies (Yates et al. \cite{yates97}). Interstellar and stellar 22-GHz
masers require relatively high H$_2$ ($<10^8-10^{10}$~cm$^{-3}$) and water 
($10^3-10^5$~cm$^{-3}$) densities, but with an abundance ratio
to H$_2$ $<10^{-4}$, and kinetic temperatures within $200-2000$~K (Yates et al. \cite{yates97}).  

Jupiter-like planets whose atmosphere contains $\sim$80\% of H$_2$, might offer suitable conditions 
for water masers, assuming that a large amount of water vapour is present.
In Jupiter, the water vapour appears to lie at high atmospheric
altitudes in the middle stratosphere (Bergin et al. \cite{bergin00} and references
therein). The H$_2$O abundance results from both internal chemistry
and external transport through the Shoemaker-Levy 9 comet, and is
stable against photolysis and conversion to CO$_2$ over
typically 50 years (Lellouch et al. \cite{lellouch02}). The average physical conditions for the
bulk of water vapour at $p<5.5$~mbar 
are $T=150$~K, $N(\rm{H_2O})=2.8\times10^{15}$~cm$^{-2}$ and the H$_2$O abundance
ratio to H$_2$ $\sim10^{-9}$. In the case of a deep impact, a large
amount of water vapour can be detected in the hot upper atmosphere
(Bjoraker et al. \cite{bjoraker96}) where $p=3$~$\mu$bar and $T>200$~K (Young \cite{young03}).
The H$_2$O abundance ratio to H$_2$ increases to 10$^{-7}$ with higher altitudes
(Fig. 2 in Bergin et al. \cite{bergin00}). 

Another vital parameter for maser action is the
pumping efficiency. Collisional pumping with H$_2$ molecules might
occur in the hot upper atmosphere and deeper in the stratosphere,
where the temperature could increase to $>200$~K in shocked gas. These
elements suggest that suitable conditions for planetary water masers are probably met in
Jupiter-like planet atmospheres following a comet impact. The existence of water masers in Earth-like planets is
less likely because the Earth atmosphere is characterised by a large
column density of thermalised water vapour. 

Outside the Solar System, the collision of relatively large comets with a planet
could enhance the water vapour abundance and induce maser action if the pumping
and the velocity coherence are sufficiently effective along the path length. 
To estimate the physical conditions that produce masers with a peak intensity $\sim100$~mJy, 
we consider the case of a comet impact that would inject water vapour in the atmosphere of a 
Jupiter-like planet (140\,000 km in diameter). In this simplified model, the masing
path length is the thickness of the atmosphere in which water vapour is present.
It can vary between 300 km (a stratosphere path length) and $\sim2\times10^4$~km (the maximum 
tangential path length for a 1000-km atmosphere). A single tangential path length of $1.6\times10^4$~km 
is adopted for the purpose of this work. It corresponds to the maximum path length for a height
of 500 km in the atmosphere. The maser amplifies a background radiation 
with a brightness temperature corresponding to the local atmospheric temperature. Two kinetic H$_2$ 
temperature ($T_{\rm{H_2}}$) cases are studied: 150~K (stratosphere) and 300~K (hot upper atmosphere or shocked gas). 
The unsaturated maser brightness temperature is given by $T_{\rm{H_2}}$e$^{\tau}$,
where $\tau$ is the maser gain coefficient. The value of $\tau$ is, in first approximation, proportional to the path 
length, the velocity distribution function ($1/\Delta\nu_D$ at the line centre, where $\Delta\nu_D$ is the thermal 
doppler broadening) and the population difference ($n_1-n_2$, where $n_1$ and $ n_2$ are the population of the
lower and upper levels, respectively). Once $\tau$ is known for a given flux density translated in brightness 
temperature, one can deduce the total column density of water vapour by assuming that the pumping efficiency 
($n_1-n_2/n_1+n_2$) is $\sim1$\% and $n_1$ nearly obeys the Boltzmann distribution.

\begin{table}
\caption[]{Physical conditions to produce a 100-mJy Jovian water
maser at a distance $D$. Standard maser theory is used (Reid \& Moran \cite{reid88}). $R$ is the
radius of the comet nucleus and $M_{\rm{H_2O}}$ is the total mass of water in it. A mean
ice density of 300 kg m$^{-3}$ is used for the comet
nucleus. $N_{\rm{H_2O}}$ is the column density in the planet atmosphere.}
\begin{center}
\footnotesize
\begin{tabular}{lllll}
\hline
\hline
$T_{\rm{H_2}}$(K)             &  150                & 150              &   300   & 300 \\
$D$ (pc)                       &  10                 & 100              &   10    & 100 \\
\hline
$N_{\rm{H_2O}}$ ($10^{19}$ cm$^{-2}$) &  6.9 & 8.6 & 3.2 & 4.0 \\
$R$ (km)                       &  6.3               & 6.8              &   4.9   & 5.2  \\
$M_{\rm{H_2O}}$ ($10^{14}$ kg)  &  3.2 & 4.0 &  1.4 & 1.8 \\
\hline
\end{tabular}
\end{center}
\end{table}

Table 1 presents the physical conditions required to produce a 100-mJy maser at distances of 10-100 pc.
The resulting H$_2$O density varies between 10$^{10}$ and 10$^{11}$ cm$^{-3}$ 
depending on $T_{\rm{H_2}}$. These values do not satisfy the density conditions and longer path 
lengths ($>>10^6$ km) are desirable to avoid high density that will quench the inversion population.
If the maser occurs in the planetary atmosphere, its  H$_2$O column density has to be enhanced by a factor 
10$^3$ to 10$^5$ to allow maser action assuming initial conditions comparable to those in Jupiter (see above). 
The H$_2$O abundance ratio to H$_2$ varies between $10^{-6}$ and $10^{-4}$. 
A large amount of water ice, $10^{14}-10^{15}$ kg, is needed to fill in part of
the planet atmosphere (Table 1). In comparison, Bjoraker et al. (\cite{bjoraker96}) measured a column density of
10$^{18}$ cm$^{-2}$ and a total water mass of 10$^9$ kg toward the impacts of the Shoemaker-Levy 9 comet.

In summary, an extrasolar planet must undergo a much more severe bombardment
than Jupiter in 1994 to produce a detectable water maser at a level of 100 mJy. However, the presence of  
100-mJy masers is difficult to explain in a planetary environment in terms of standard H$_2$O maser theory 
(e.g. Reid \& Moran \cite{reid88}). Elitzur et al. (\cite{elitzur89}) have characterised the maser emission measure with a parameter
$\xi$ (see Eq. 2.1 in Elitzur et al.), which is directly proportional to the water vapour
abundance ratio to H$_2$, the square of the H$_2$ density and the path length. To compensate the relatively
short planetary path length and obtain $\xi>1$, either large H$_2$O abundance ratio or large gas density is required, 
which might be unrealistic in the first case and could quench the maser in the second case.
Finally, the velocity coherence is implicitely assumed to be reached along the full path length, which is unlikely 
to occur along a sufficiently large path length in such a turbulent atmosphere.

\section{Observations}

Interferometers allow both high angular resolution and high
sensitivity observations. These technical characteristics are ideal in
the search for water masers toward extrasolar planets. A small beam
could clearly establish the connection between water masers and exoplanets in
case of detection. Furthermore, masers from a planetary
atmosphere are probably highly variable as the planet rotates (only 10-hour sidereal period for Jovian planets).

In April 2003, we used the newly upgraded ATCA system at 12 mm in the EW352-baseline configuration plus antenna 6. 
Eighteen extrasolar planets were searched for 22-GHz water masers down to a
noise rms of about 25 mJy~beam$^{-1}$ (average weather conditions), excluding antenna 6 in the
rms estimate.  A bandwidth of 16 MHz and 512 spectral channels were used, 
which give a spectral resolution of 31 kHz or 0.4 km~s$^{-1}$ and 
a total velocity range of 216 km~s$^{-1}$ centred at $V_{lsr}=0$~km~s$^{-1}$. An angular resolution 
of about 10 arcsec was obtained with the EW352 configuration,
allowing us to probe the inner $\sim30-1200$~AU of the planetary systems depending on their
distance to us. 
Each target was observed in a straight 30-minute scan in average, preceded and
followed by a 3-min scan on the phase calibrator. The high 
sensitivity of ATCA at 12 mm allows us to reach a 4-$\sigma$ rms of
100 mJy~beam$^{-1}$ in 30 minutes, which is $1/20$ of the Jupiter sidereal
period. 

\begin{table*}
\caption[]{Exoplanet sample and results. The age of each star is estimated with log($t$)=$10.725-1.334{\times}R+0.4085{\times}R^2-0.0522{\times}R^3$
(Donahue \cite{donahue93}), where $R$ is the chromospheric activity times $10^5$. Notes: $^1$Tau1 Gru; $^2$Iot Hor; $^3$Eps Eri; $^4$Eps Ret.}
\begin{center}
\footnotesize
\begin{tabular}{llllllllll}
\hline
\hline
 HD   &  Spectral  & Mass    & Distance & Age & \multicolumn{2}{c}{Coordinates} & Radial vel. & rms & Detection \\ 
\cline{6-7}
 name & type       & (M$_{\odot}$) & (pc) & (10$^9$yr) & RA(J2000)  &    Dec(J2000) & (km s$^{-1}$) & (mJy) &  \\
\hline
HD 169830 & F8V &	1.40 &  36.32 &  4.3 & 18:27:49.5 & -29:49:00 &  ? & 28 & No \\
HD 179949	& F8V & 1.24 &   27.05  & 2.0 & 	 19:15:33.2  & -24:10:45 & -25.5 & 24 & No \\
HD 202206& G6V& 	1.15 &   46.34 &  ? 	   &       21:14:57.8& -20:47:21 & 14.7 & 26 & No \\
HD 213240& G4IV& 1.22&    40.75 &  2.7 & 	 22:31:00.4 & -49:25:59 & -0.9 & 26 & No  \\        
HD 216435$^1$ &  G0V& 1.25 &   33.29 & 5.6 & 	 22:53:37.9 &  -48:35:53 & -1.0 & 26 & No \\
HD 142 & 	G1IV& 	1.10 &   25.64  & 4.2 & 	 00:06:19.2& -49:04:30 & 2.6 & 25 & No \\
HD 2039 &   G2.5IV-V& 0.98 &   89.85 & 4.0 & 00:24:20.3 & -56:39:00 & ? & 27  & No \\
HD 6434& G3V & 1.00 &   40.32 &  3.7 & 	 01:04:40.2 &  -39:29:17 & 22.4 & 27 & No \\
HD 13445& K0V& 0.80 &   10.91 & 2.2 & 	 02:10:25.9 &  -50:49:25 & 53.1 &  26 & No \\
HD 17051$^2$ & G0V& 1.03 &   17.24 & 1.6 & 	 02:42:33.5 &  -50:48:01& 15.0 & 25 & No \\
HD 22049$^3$ & K2V& 0.80 &   3.22  & 0.7 & 	 03:32:55.8 &  -09:27:29 & 15.5 & 26 & No \\
HD 23079& F9V& 1.10 &   34.60 & 4.8 & 	 03:39:43.1 &  -52:54:57 & ? & 29 & No \\
HD 27442$^4$ & K2IV& 1.20 &   18.23 & ? &         04:16:29.0 & -59:18:07 & 29.3 & 29 & No \\
HD 30177& G8V& 0.95 &   54.71 &  7.4 & 	 04:41:54.4 & -58:01:14 & ? & 30  & No \\
HD 47536& K0III& 1.10 &   121.36 & $>5$	&  06:37:47.6 &  -32:20:23 & 78.8 & 31 & No? \\
HD 73526& G7V& 1.02 &   94.61  &  ? 	&  08:37:16.5 &  -41:19:08 & ? & 29 & No \\
HD 75289& G0V& 1.15 &   28.94  & 4.9 & 	 08:47:40.4 & -41:44:12 & 14.0 & 27 & No \\
HD 83443& K0V& 0.79 &   43.54  & 3.2 & 	 09:37:11.8 & -43:16:19 & 27.6 & 29 & No \\
\hline
\end{tabular}
\end{center}
\end{table*}

\section{Data analysis, results and discussion}

The data were analysed with {\it Miriad} in an essentially identical
method as for the centimetre data analysis (Sault \& Killeen \cite{sault04}). The amplitude calibration was achieved on Uranus. The bandpass
solutions were obtained with a strong calibrator, 1730-130. Pointing
corrections were estimated and applied every 2 hours in average. The
phase calibration was made with secondary calibrators. 
A couple of well known
maser sources were successfully observed during the experiment to check the new 12-mm system.

Vector-averaged cross-correlated spectra were
generated from the data cubes for each extrasolar planet. All
baseline-averaged spectra were first inspected. 

The individual baseline spectra
were also averaged over various time ranges ($\sim2-20$ min) and searched for artificial lines 
and peaks. Possible maser lines from a rotating planetary atmosphere could then be monitored 
in velocity as long as the maser was bright enough to be identified in a few minutes. Conversely,
non detection could be the result of signal dilution in velocity/channels over a significative 
fraction of sidereal period. If a signal was visualised, the spectrum was Hanning smoothed and 
checked again. Both polarisation spectra were also inspected in a similar fashion.
The 1-$\sigma$ rms on the noise baseline varies between 24 and 31 mJy for our source 
sample (Table 2). This includes the noise bias value,
which is a residual noise level obtained after vector averaging. After all the above
checks, a signal was considered as a possible detection when the flux intensity 
was greater than $\sim4$ $\sigma$ and the line was still visible after Hanning smoothing.

The vector-averaged spectrum for HD\,47536, a KIII giant star with a giant planet candidate 
(Setiawan et al. \cite{setiawan03}), exhibits a noise level greater than $\sim3-4$ $\sigma$ in a few 
channels. The intensity level reaches 100~mJy around $V_{lsr}=-36$~km~s$^{-1}$. 
Channel maps were produced
after averaging over 7 channels ($3$~km~s$^{-1}$) that corresponds to the {\it FWHM} of the putative maser line. A 
100-mJy signal was identified in the channel map 
at the averaged velocity of $-36$~km~s$^{-1}$. However, a few channel maps exhibit a flux density
level up to 50 mJy. It is thus possible that the emission at $-36$~km~s$^{-1}$ is associated with a noise peak
rather than a true signal. New ATCA observations in 2004 failed to confirm this possible detection.

Finally, no maser is observed toward Epsilon Eridani, a target that was reported to exhibit a possible
maser line (Cosmovici et al. \cite{cosmovici02}). This result confirms other non-detection reports by Greenhill (IAU circular 7985)
and Butler et al. (\cite{butler02}). Our sensitivity limit for Epsilon Eridani was 26 mJy~beam$^{-1}$. This value
compares well with those achieved by other instruments such as the VLA and large aperture single dish telescope. 
The possible signal toward Eps Eri was measured at an intensity level $>3\sigma$ (with $1\sigma=200$~mJy) with 
the Medicina 32-m radio telescope (Cosmovici, priv. comm.). The ATCA non-detection results suggest that the
lines observed by Cosmovici et al. are (i) transient masers, (ii) signals from another location in the 
sky that was detected in the large beam ($\sim1.5'$) of the Medicina 32-m telescope or 
(iii) artificial lines (e.g. RFI). 

In conclusion, our observations have confirmed that detectable water masers in planetary systems
are rare phenomena. Whether these phenomena can
occur in an exoplanetary environment is an issue that can only be solved with very high-sensitivity
radio telescopes. Our results and conclusions are comparable to an unsuccessful search for cyclotron 
masers toward extrasolar systems (Bastian et al. \cite{bastian00}). Randomly distributed observations in time with 
large aperture radio instruments (e.g. SKA in the future) might possibly be useful to further address this issue.

\begin{acknowledgements}
The Australia Telescope Compact Array is part of the Australia Telescope which is funded by the Commonwealth 
of Australia for operation as a National Facility managed by CSIRO. We thank Ray Norris, Bob Sault and Cormac
Purcell for their help in the ATCA observations.
\end{acknowledgements}

{}


\begin{thebibliography}{}
\bibitem[2000]{bastian00} 
Bastian, T.S., Dulk, G.A., \& Leblanc, Y. 2000, ApJ, 545, 1058 
\bibitem[2000]{bergin00}
Bergin, E.A., Lellouch, E., Harwit, M., et al. 2000, ApJ, 539, L147
\bibitem[1996]{bjoraker96}
Bjoraker, G.L., Stolovy, S.R., Herter, T.L., Gull, G.E., \& Pirger, B.E. 1996, Icarus, 121, 411
\bibitem[2003]{boonman03}
Boonman, A.M.S., Doty, S.D., van Dishoeck, E.F., et al. 2003, A\&A, 406, 937
\bibitem[1989]{brown89}
Brown, J.A., Johnson, H.R., Cutright, L.C., Alexander, D.R., \& Sharp, C.M. 1989, ApJS, 71, 623
\bibitem[2002]{butler02}
Butler B.J., Chandler, C.J., Claussen, M.J., \& Greenhill, L.J. 2002, AAS, 201, 4612
\bibitem[1990]{cernicharo90}
Cernicharo, J., Thum, C., Hein, H., et al. 1990, A\&A, 231, 15
\bibitem[1969]{cheung69}
Cheung, A.C., Cudaback, D.D., Rank, D.M., et al. 1969, BAAS, 1, 236
\bibitem[1990]{chyba90}
Chyba, C.F. 1990, Nature, 343, 129 
\bibitem[1984]{claussen84}
Claussen, M.J., Heiligman, G.M., \& Lo, K.Y. 1984, Nature, 310, 298 
\bibitem[1996]{cosmovici96}
Cosmovici, C.B., Montebugnoli, S., Orfei, A. Pogrebenko, S., \& Colom, P. 1996, P\&SS, 44, 735
\bibitem[2002]{cosmovici02}
Cosmovici, C., Teodorani, M, Montebugnoli, S, \& Maccaferri, G. 2002, presentation at the Second
European Workshop on Exo/Astrobiology, Graz, Austria
\bibitem[1990]{deguchi90}
Deguchi, S., \& Nguyen-Q-Rieu. 1990, ApJ, 360, 27
\bibitem[1976]{dickinson76}
Dickinson, D.F. 1976, ApJSS, 30, 259
\bibitem[1993]{donahue93}
Donahue, R.A. 1993, PhD thesis, New Mexico State University
\bibitem[1989]{elitzur89}
Elitzur, M., Hollenbach, D.J., \& McKee, C.F. 1989, ApJ, 346, 983
\bibitem[2002]{engels02}
Engels, D. 2002, A\&A, 388, 252
\bibitem[1979]{genzel79}
Genzel, R., \& Downes, D. 1979, A\&A, 72, 234
\bibitem[2003]{lecacheux03}
Lecacheux, A., Biver, N., Crovisier, J., et al. 2003, A\&A, 402, 55
\bibitem[2002]{lellouch02}
Lellouch, E., B\'ezard, B., Moses, J.I., et al. 2002, Icar, 159, 112
\bibitem[1990a]{menten90a}
Menten, K.M., Melnick, G.J., \& Phillips, T.G. 1990, ApJ, 350, 41
\bibitem[1990b]{menten90b}
Menten, K.M., Melnick, G.J., Phillips, T.G., \& Neufeld, D.A. 1990, ApJ, 363, 27
\bibitem[1988]{reid88}
Reid, M.J., \& Moran, J.M. 1998, in Galactic and extragalactic radio astronomy (2nd edition), Berlin and New York, Springer-Verlag, 255
\bibitem[2004]{sault04}
Sault, B., \& Killeen, N. 2004, The Miriad User Guide, http://www.atnf.csiro.au/computing/software/miriad/
\bibitem[2003]{setiawan03}
Setiawan, J., Hatzes, A.P., von der Luhe, O., et al., 2003, A\&A, 398, L19
\bibitem[1983]{shoemaker83}
Shoemaker, E.M. 1983, AREPS, 11, 461
\bibitem[2001]{takaba01}
Takaba, H., Iwate, T., Miyaji, T., \& Deguchi, S. 2001, PASJ, 53, 517
\bibitem[1998]{torrelles98}
Torrelles, J.M., G{\'o}mez, J.F., Rodr{\'i}guez, L.F., et al. 1998, ApJ, 505, 756
\bibitem[1997]{tsuji97}
Tsuji, T., Ohnaka, K., Aoki, W., \& Yamamura, I. 1997, A\&A, 320, 1
\bibitem[1995]{wallace95}
Wallace, L., Bernath, P., Livingston, W., et al. 1995, Sci, 268, 1155 
\bibitem[1980]{waters80}
Waters, J.W., Kakar, R.K., Kuiper, T.B.H., et al. 1980, ApJ, 235, 57
\bibitem[1997]{yates97}
Yates J.A., Field, D., \& Gray, M.D. 1997, MNRAS, 285, 303
\bibitem[2003]{young03}
Young, R.E. 2003, NewAR, 47, 1
\end{thebibliography}
\end{document}